\documentclass[prd,onecolumn,notitlepage,amsmath,nofootinbib,superscriptaddress,showpacs,showkeys]{revtex4-1}

\usepackage[brazil,english]{babel}
\usepackage{amsmath}
\usepackage{amssymb}

\usepackage{ulem}
\PassOptionsToPackage{normalem}{ulem}

\usepackage{amsfonts}
\usepackage{graphicx}



\begin{document}

\title{A Hamiltonian approach for the Thermodynamics of AdS black holes}

\author{M. C. Baldiotti}
\email{baldiotti@uel.br}
\affiliation{Departamento de F\'{\i}sica, Universidade Estadual de Londrina, 86051-990,
Londrina-PR, Brazil.}

\author{R. Fresneda}
\email{rodrigo.fresneda@ufabc.edu.br}
\affiliation{Universidade Federal do ABC, Av. dos Estados 5001, 09210-580, Santo
Andr\'{e}-SP, Brazil}

\author{C. Molina}
\email{cmolina@usp.br}
\affiliation{Escola de Artes, Ci\^{e}ncias e Humanidades, Universidade de S\~{a}o Paulo \\ 
Av. Arlindo Bettio 1000, CEP 03828-000, S\~{a}o Paulo-SP, Brazil}

\begin{abstract}
In this work we study the Thermodynamics of $D$-dimensional 
Schwarzschild-anti de Sitter (SAdS) black holes. The minimal 
Thermodynamics of the SAdS
spacetime is briefly discussed, highlighting some of its strong points
and shortcomings. The minimal SAdS Thermodynamics is extended within
a Hamiltonian approach, by means of the introduction of an additional
degree of freedom. We demonstrate that the cosmological constant can
be introduced in the thermodynamic description of the SAdS black hole
with a canonical transformation of the Schwarzschild problem, closely
related to the introduction of an anti-de Sitter thermodynamic volume.
The treatment presented is consistent, in the sense that it is compatible
with the introduction of new thermodynamic potentials, and respects
the laws of black hole Thermodynamics. By demanding homogeneity of
the thermodynamic variables, we are able to construct a new equation
of state that completely characterizes the Thermodynamics of SAdS
black holes. The treatment naturally generates phenomenological constants
that can be associated with different boundary conditions in underlying
microscopic theories. A whole new set of phenomena can be expected
from the proposed generalization of SAdS Thermodynamics. 
\end{abstract}


\maketitle

\section{Introduction}

The emergence of the anti-de Sitter (AdS)/conformal field theory (CFT)
correspondence \cite{maldacena,witten,gubser} is a milestone of
contemporary high energy physics. In broad terms, the AdS/CFT relation
postulates a dictionary associating gravity physics in an asymptotically
AdS geometry with field theory in the boundary of the AdS space. Although
having originated in a context involving string theory, extended gauge/gravity
dualities have been suggested, with applications in a variety of physical
scenarios, from fundamental quantum gravity models to phenomenological
condensed matter systems. Still, in many ways, the understanding of
the AdS/CFT dictionaries is a work in progress. Paramount to this
endeavor is the proper characterization of the semi-classical and
quantum physics of the asymptotically AdS spacetimes.

We consider thermodynamic aspects of asymptotically AdS spacetimes,
a theme that has received considerable attention in recent years \cite{brown,louko,hemming,grumiller,rajeev,morgan,miranda,hubeny,Rabin,menoufi,cardoso,myung,lemos}.
More specifically, we focus on the $D$-dimensional Schwarzschild-anti
de Sitter (SAdS) black hole. We consider a macroscopic description,
where the black hole is in equilibrium with a thermal atmosphere generated
by Hawking emission \cite{wald}. In the usual thermodynamic setup
for the SAdS geometries, denoted here as the minimal SAdS Thermodynamics,
there is only one thermodynamic variable. In this minimal thermodynamic
approach the first law of Thermodynamics is not consistent with the
Smarr formula \cite{kastor2009}. Therefore, the minimal description
is not a stricto sensu Thermodynamics.

Moreover, if the number of degrees of freedom in the thermodynamic
description of the SAdS black hole is increased, one obtains a Thermodynamics
which is closer to the standard Thermodynamics associated to matter
systems \cite{Joh14}. The underlying reason for this result is that
no usual thermodynamic system has only one degree of freedom. In fact,
in a setup with one degree of freedom it is not possible to establish
a distinction between isothermic and isentropic processes, and the very
notion of thermodynamic temperature is not well-defined. Besides,
AdS/CFT duality suggests that SAdS gravity is associated to a finite-temperature
$\mathcal{N}=4$ supersymmetric Yang--Mills theory \cite{nat2015}.
Since the field theory side of the correspondence is compatible with
a proper thermodynamic treatment, it is natural that a standard Thermodynamics
should be present at the gravity side. The basic question here is
how the minimal SAdS Thermodynamics can be extended in order to have
a proper Thermodynamics.

\newpage

The idea of extending the usual black hole Thermodynamics with the
introduction of the cosmological constant $\Lambda$ as a thermodynamic
variable was proposed in \cite{21,seki2006,29}. However, the physical
interpretation of the conjugate variable associated to $\Lambda$
is still an open problem. For instance, some authors proposed that
this conjugate variable could be a thermodynamic volume \cite{kastor2009,Dol1,2011}.
In this proposal, $\Lambda$ would be interpreted as pressure, and
the black hole mass would be identified with the enthalpy, and not
with the internal energy (as is usually the case). The drawback here
is that, if the black hole has no electric charge or spin, the interpretation
of $\Lambda$ as pressure leads to inconsistencies when other thermodynamic
potentials are considered \cite{seki2006,Dol1,KubMan2012}.

We demonstrate that the cosmological constant can be introduced in
the thermodynamic description of the SAdS black hole, though not as
a new thermodynamic variable (as proposed for example in \cite{21,seki2006,29}),
but as a function on phase space through a new equation of state.
This new equation of state is closely related to a generalized thermodynamic
volume. To this aim, we use the Hamiltonian approach to Thermodynamics
developed in \cite{baldfresmol2016}. A thermodynamic description
based on symplectic geometry is introduced, and thermodynamic equations
of state are realized as constraints on phase space. Since it is a
symplectic formalism, the necessary addition of degrees of freedom
in order to generalize the minimal SAdS Thermodynamics is achieved
in a natural manner by enlarging phase space. The method helps
one to explore consequences of having additional degrees of freedom in the SAdS Thermodynamics.

The thermodynamic treatment presented here is consistent, in the sense
that it is compatible with the introduction of new thermodynamic potentials,
and respects the laws of black hole Thermodynamics. By demanding homogeneity
of the thermodynamic variables, we are able to construct a new equation
of state that completely characterizes the Thermodynamics of SAdS
black holes. The treatment naturally generates phenomenological constants
that can be associated with different boundary conditions in underlying
microscopic theories. A whole new set of phenomena can be expected
from the proposed generalization of SAdS Thermodynamics.

The structure of this work is as follows. In section~\ref{sec:AdS-black-holes}
we review the basic results concerning the $D$-dimensional SAdS spacetime
and the usual (and the not quite usual) thermodynamic treatments to
the SAdS black hole. In section~\ref{sec:A-Symplectic-approach}
the symplectic approach is used for the characterization of the SAdS
Thermodynamics. In section~\ref{sec:The-Euler-relation} the additional
condition of homogeneity of the thermodynamic variables is imposed.
A complete macroscopic scenario is developed in the process. Final
considerations and some perspectives of future developments are presented
in section~\ref{sec:Conclusions}. Throughout this paper, we use
signature $(-++ \cdots +)$ and natural units with $G=\hbar=c=k_{B}=1$.

\section{\label{sec:AdS-black-holes}Schwarzschild-A\MakeLowercase{d}S black holes and Thermodynamics}

\subsection{Elements of the SAdS geometry}

We study the Thermodynamics of asymptotically AdS black holes in equilibrium
with a thermal atmosphere generated by Hawking emission \cite{wald}.
The approach here is purely macroscopic, with quantum effects taken
into account only implicitly.

The family of static black holes are modeled by the $D$-dimensional
SAdS spacetime.%
\footnote{Perhaps this geometry should be denoted as the ``Tangherlini anti-de
Sitter spacetime'', since it is the generalization of the Tangherlini
metric \cite{tangherlini}. However, following the usual terminology, 
we will call the geometry ``Schwarzschild-anti de Sitter''.%
} 
The SAdS metric can be seen
as the spherically symmetric solution of the $D$-dimensional Einstein's
equations assuming vacuum and a negative cosmological constant $\Lambda$
\cite{nat2015,Boucher}.

For a spherically symmetric and asymptotically anti-de Sitter black
hole, the associated line element is 
\begin{equation}
ds^{2}=g_{tt}(r)\, dt^{2}+g_{rr}(r)\, dr^{2}+r^{2}d\Omega_{D-2}^{2}\,,\label{metric}
\end{equation}
where $d\Omega_{D-2}^{2}$ is the line element of the unit sphere
$S^{D-2}$, 
\begin{equation}
d\Omega_{D-2}^{2}=\left(d\theta_{1}\right)^{2}+\sin^{2}\theta_{1}\,\left(d\theta_{2}\right)^{2}+\cdots+\sin^{2}\theta_{1}\cdots\,\sin^{2}\theta_{D-2}\,\left(d\theta_{D-2}\right)^{2}\,.\label{dOmega2}
\end{equation}
The coordinates $\{t,r\}$ refer to $\mathcal{M}^{2}$, and the coordinate
system based on \linebreak{}
 $\left\{ \theta_{1},\theta_{2},\ldots,\theta_{D-2}\right\} $ describes
$S^{D-2}$. The metric functions $g_{tt}(r)$ and $g_{rr}(r)$ are
given by 
\begin{equation}
-g_{tt}\left(r\right)=\frac{1}{g_{rr}(r)}=1-\frac{\tilde{M}}{r^{D-3}}-\tilde{\Lambda}r^{2}\,.\label{h}
\end{equation}
The constants $\tilde{M}$ and $\tilde{\Lambda}$ in Eq.~(\ref{h})
are expressed in terms of the mass parameter $M$ and the cosmological
constant $\Lambda$ as 
\begin{equation}
\tilde{M}=\frac{16\pi}{\left(D-2\right)B_{D-2}}\, M\,,\,\, B_{D-2}=\frac{2\pi^{\frac{D-1}{2}}}{\Gamma\left(\frac{D-1}{2}\right)}\,,\,\,\tilde{\Lambda}=\frac{2}{\left(D-1\right)\left(D-2\right)}\,\Lambda\,.\label{h-constants}
\end{equation}
In Eq.~(\ref{h-constants}), $B_{D-2}$ is the ``area'' of $S^{D-2}$,
that is, the canonical volume associated with the induced metric given
by $d\Omega_{D-2}^{2}$ in Eq.~(\ref{dOmega2}).

The zeros of the function $g_{tt}(r)$ in Eq.~(\ref{h}) determine
the causal structure of the SAdS spacetime. If $M>0$ and $\Lambda<0$,
$g_{tt}$ has a unique positive real zero, denoted as $r_{+}$. The
parameter $\tilde{M}$ is associated with $r_{+}$ and $\tilde{\Lambda}$
by the relation 
\begin{equation}
\tilde{M}=r_{+}^{D-3}\left(1-\tilde{\Lambda}r_{+}^{2}\right)\ .\label{mass-1}
\end{equation}
The coordinate system $\left(t,r,\theta_{1},\ldots,\theta_{D-2}\right)$
is valid only for $r_{+}<r<\infty$. But the geometry can be extended
by the usual methods, and in its maximal extension the hypersurface
$r=r_{+}$ is a Killing horizon, with a surface gravity given by 
\begin{equation}
\kappa=\lim_{r\rightarrow r_{+}}\,\frac{1}{2}\frac{d}{dr}\left(\sqrt{-\frac{g_{tt}(r)}{g_{rr}(r)}}\right)=\frac{1}{2}\left[\frac{D-3}{r_{+}}-\left(D-1\right)\tilde{\Lambda}r_{+}\right] \,,
\label{kappa-1}
\end{equation}
where Eq.~(\ref{mass-1}) was used. The Killing horizon area $A$
is written in terms of $r_{+}$ as 
\begin{equation}
A=r_{+}^{D-2}\, B_{D-2}\,.\label{area}
\end{equation}

\subsection{\label{sub:Minimal-Thermodynamics-for}Minimal Thermodynamics for
the SAdS black holes}

The usual thermodynamic interpretation to the Schwarzschild-AdS spacetime
postulates an ensemble of AdS black holes with no charge and no angular
momentum, each one in equilibrium with its thermal Hawking atmosphere
\cite{wald,nat2015}. In the minimal thermodynamic setup for the
SAdS geometries, there is only one thermodynamic variable, the entropy
$S$, and therefore the fundamental equation has the form $U=U\left(S\right)$.
In this setup, the internal energy $U$, entropy $S$ and thermodynamic
temperature $T$ are defined as 
\begin{equation}
U\equiv M\,,\,\, S\equiv\frac{A}{4}\,,\,\, T\equiv\frac{\kappa}{2\pi}\,.\label{usual}
\end{equation}
With the relations~(\ref{usual}) and the results presented in the
previous section, it is straightforward to verify that 
\begin{equation}
\frac{\partial U}{\partial S}=T\,,\label{duds}
\end{equation}
and therefore 
\begin{equation}
dM=\frac{\kappa}{8\pi}\, dA\Rightarrow dU=TdS\,.\label{dm}
\end{equation}

Some useful relations between the several constants which characterize
the $D$-dimensional SAdS geometry are given in the following. From
Eqs.~(\ref{mass-1}) and (\ref{area}), it follows that 
\begin{equation}
M=\left(D-2\right)\frac{B_{D-2}}{16\pi}\left(\frac{A}{B_{D-2}}\right)^{\frac{D-3}{D-2}}\left[1-\tilde{\Lambda}\left(\frac{A}{B_{D-2}}\right)^{\frac{2}{D-2}}\right]\,.\label{eq:M}
\end{equation}
Moreover, using expressions~(\ref{kappa-1}) and (\ref{area}), one
has 
\begin{equation}
2\kappa=\left(D-3\right)\left(\frac{B_{D-2}}{A}\right)^{\frac{1}{D-2}}-\left(D-1\right)\tilde{\Lambda}\left(\frac{A}{B_{D-2}}\right)^{\frac{1}{D-2}}\,.\label{eq:kappa}
\end{equation}
Combining Eqs.~(\ref{eq:M}) and (\ref{eq:kappa}), we obtain a relation
where the cosmological constant $\Lambda$ does not explicitly appear:
\begin{equation}
8\pi M\frac{D-1}{D-2}=A\,\left(\frac{B_{D-2}}{A}\right)^{\frac{1}{D-2}}+\kappa A\,.\label{eq_no_Lambda}
\end{equation}

Rewriting the relation~(\ref{eq_no_Lambda}) in terms of the thermodynamic
quantities, we obtain the equation of state 
\begin{equation}
\frac{D-1}{D-2}U-TS=\frac{1}{2\pi}\left(\frac{B_{D-2}}{4}\, S^{D-3}\right)^{\frac{1}{D-2}}\,.\label{eq:kappa2}
\end{equation}
 This is the only equation of state of the thermodynamic description
of the SAdS ensemble.

The minimal approach for the SAdS Thermodynamics thus sketched is
simple and consistent, as far as the thermodynamic potentials are
concerned. However, it is rather limited: there is only one thermodynamic
degree of freedom. More importantly, the theory is not homogeneous.
Since homogeneity is required for extensivity \cite{kastor2009},
as well as for the existence of an integrating factor for the reversible
heat exchange \cite{bel2-2005}, the minimal AdS Thermodynamics is
not a stricto sensu Thermodynamics. In the following, we will briefly
review the first attempt to introduce more structure to this model.

\subsection{\label{first_extension}Attempting to generalize the minimal SAdS
Thermodynamics}

In this subsection we prelude the main development of this
work, reviewing a first attempt to generalize the minimal AdS Thermodynamic. 
It is an insightful approach, although not entirely consistent, as we will see in the following.

The starting point in the extension of the minimal AdS black hole
Thermodynamics is to consider the cosmological constant $\Lambda$
as a thermodynamic variable \cite{21,seki2006,29}. The internal
energy $U$ in the minimal AdS Thermodynamics (that is, the mass parameter
$M$) is not a first order homogeneous function. Therefore, the Euler
relation does not have the usual form. If $\Lambda=0$ and $D=4$
it can be written as $M=2TS$, and it is known as the Smarr formula
\cite{smarr73}. The techniques introduced in \cite{kastor2009,seki2006}
extend the Smarr formula to the $D$-dimensional SAdS scenario. It
follows that, for the case of the AdS black hole with no charge or
angular momentum, one has the $D$-dimensional Smarr formula \cite{kastor2010},
\begin{equation}
\left(D-3\right)M=\left(D-2\right)\frac{\kappa A}{8\pi}-2\frac{\theta\Lambda}{8\pi}\,\,.\label{smarr}
\end{equation}
In expression~(\ref{smarr}), $\theta$ represents a new thermodynamic
variable, conjugated to $\Lambda$. Using Eqs.~(\ref{eq:kappa}),
(\ref{eq_no_Lambda}) and the Smarr formula~(\ref{smarr}) for $D$
dimensions, one obtains 
\begin{equation}
\theta=-\frac{B_{D-2}}{\left(D-1\right)}\left(\frac{4S}{B_{D-2}}\right)^{\frac{D-1}{D-2}}=-\frac{B_{D-2}}{\left(D-1\right)}r_{+}^{D-1}\,.\label{vol}
\end{equation}

Result~(\ref{vol}) identifies $\theta$ as the ``volume extracted
from spacetime'' by the black hole \cite{kastor2009}. This identification
suggests that the cosmological constant $\Lambda$ should be interpreted
as a pressure. Also, from this point of view, the black hole mass
$M$ should not be identified with its internal energy, but with its
enthalpy $H$ \cite{kastor2009}. 
In this generalized Thermodynamics of the AdS black hole, there are two independent thermodynamic variables, such that 
\begin{equation}
H\equiv M\,,\,\, S\equiv\frac{A}{4}\,,\,\, T\equiv\frac{\kappa}{2\pi}\,,\,\, P\equiv-\frac{\Lambda}{8\pi}\,,\,\, V\equiv-\theta\,,\label{new_thermo}
\end{equation}
where $P$ is the thermodynamic pressure associated to $V$.

However, the definition of the enthalpy as the parameter $M$ leads
to an inconsistency in the thermodynamic description. In fact, it
follows from the definitions in Eq.~(\ref{new_thermo}) that 
\begin{equation}
\left.\frac{\partial H}{\partial P}\right|_{S}=V\,.
\end{equation}
The new internal energy $U$ in the proposed generalized Thermodynamics
can now be determined. Using Eqs.~(\ref{smarr}) and (\ref{eq_no_Lambda}),
\begin{equation}
U=H-PV=\left(D-2\right)\frac{\, B_{D-2}}{16\pi}\left(\frac{4S}{B_{D-2}}\right)^{\frac{D-3}{D-2}}\,.\label{u}
\end{equation}
As a result, Eq.~(\ref{u}) does not give Eq.~(\ref{duds}), that
is, 
\begin{equation}
\frac{\kappa}{2\pi}=T\ne\frac{\partial U}{\partial S}\,,
\end{equation}
and hence the Thermodynamics defined by Eq.~(\ref{new_thermo}) is
not consistent. Indeed, this problem is a consequence of the singularity
of the Legendre transformation of the pair $\left(P,V\right)$, due
to the fact that $V$ does not depend on $P$ \cite{2011}.

In summary, this approach is not consistent, since the definition of internal energy is not compatible with the definition of temperature. Although here one can argue in favor of a thermodynamic volume, a more robust theoretical framework is necessary. This is done in the next sections.

\section{\label{sec:A-Symplectic-approach}A Hamiltonian approach to Thermodynamics
of Schwarzschild-A\MakeLowercase{d}S \\ black holes}

\subsection{Extended SAdS Thermodynamics}

In the Hamiltonian approach to Thermodynamics \cite{baldfresmol2016},
one realizes all equations of state of a thermodynamic system as constraints
on phase space. Given a thermodynamic potential $M$, its differential
$dM$ is given by the canonical tautological form $pdq$ on the constraint
surface. One can extend the phase space by the addition of a canonical
pair $\left(\xi,\tau\right)$ such that the tautological form $pdq+\xi d\tau$
reduces to the Poincar\'e-Cartan form $pdq-hd\tau$ on the constraint
surface $H=\xi+h\left(q,p,\tau\right)$. Thus, one obtains a description
in the reduced phase-space $\left(p,q\right)$ as well as in the extended
phase-space $\left(p,q;\xi,\tau\right)$. In this way, all thermodynamic
potentials are seen to be related by canonical transformations, giving
equivalent representations. In other words, one is able to increase
the degrees of freedom of a thermodynamic analog.

The minimal SAdS Thermodynamics has only one free variable, the entropy. This is a one-dimensional system and the proposed
mechanical analog will accordingly be one-dimensional. As a result,
a natural identification between mechanical and thermodynamic variables,
up to canonical transformations, is 
\begin{equation}
q=\frac{4S}{B_{D-2}}\,,\,\, p=\pi T=\frac{\kappa}{2}\,.\label{pq}
\end{equation}
The next step in creating the mechanical analog is realizing the equation
of state~(\ref{eq:kappa}), 
\begin{equation}
4p=\left(D-3\right)q^{-\frac{1}{D-2}}-\frac{2\Lambda}{\left(D-2\right)}q^{\frac{1}{D-2}}~,\label{eq:momento}
\end{equation}
as a constraint. For future reference, we rewrite~(\ref{eq:M}) using
the mechanical variables defined in Eq.~(\ref{pq}): 
\begin{equation}
M=\left(D-2\right)\frac{B_{D-2}}{16\pi}q^{\frac{D-3}{D-2}}\left[1-\frac{2\Lambda q^{\frac{2}{D-2}}}{\left(D-1\right)\left(D-2\right)}\right]~.\label{eq:Mpq}
\end{equation}

Differentiating expression~(\ref{eq:Mpq}) and using Eq.~(\ref{eq:momento}),
we arrive at 
\begin{equation}
dM=\frac{B_{D-2}}{4\pi}\left[pdq-\frac{1}{2}\frac{1}{D-1}q^{\frac{D-1}{D-2}}d\Lambda\right]\,.\label{dm2}
\end{equation}
For $\Lambda$ a function of the coordinate $q$, $\Lambda=\Lambda\left(q\right)$,
the previous expression can be written as 
\begin{equation}
dM=\varpi dq\,,\label{eq:reduced-tautological-form}
\end{equation}
where 
\begin{equation}
\varpi=\frac{B_{D-2}}{4\pi}\left[p-\frac{1}{2\left(D-1\right)}q^{\frac{D-1}{D-2}}\frac{\partial\Lambda}{\partial q}\right]\,.\label{eq:varpi}
\end{equation}
The one-form $dM$ in Eq.~(\ref{eq:reduced-tautological-form}) is
the tautological form $\alpha=\varpi dq$ written in a natural set
of coordinates and restricted to the constraint surface given by condition~(\ref{eq:momento}),
$dM=\left.\alpha\right|_{\phi=0}$, where 
\begin{equation}
\phi=p-\frac{1}{4}\left(D-3\right)q^{-\frac{1}{D-2}}+\frac{1}{2}\frac{\Lambda}{\left(D-2\right)}q^{\frac{1}{D-2}}\,.\label{eq:phi-constraint}
\end{equation}

To make this statement more precise, we introduce the symplectic form
$\omega$, 
\begin{equation}
\omega=\left(\frac{\partial\varpi}{\partial p}\right)dq\wedge dp=\frac{B_{D-2}}{4\pi}dq\wedge dp\,,\label{eq:reduced-symplectic}
\end{equation}
so that the transformation $\left(p,q\right)\mapsto\left(\varpi,q\right)$
is canonical. The description just given is the reduced phase space
detailed in \cite{baldfresmol2016}. There is a straightforward way
in which one can extend this description, which amounts to introducing
a new pair of canonical variables, $\left(\xi,\tau\right)$ such that
the expression for $dM$ becomes 
\begin{equation}
dM=\varpi dq+\xi d\tau\,.\label{eq:extended-tautological-form}
\end{equation}

From the outset, $\Lambda$ is a function on phase space, so it is
a priori a function of the coordinates $\eta=\left(\varpi,q;\xi,\tau\right)$.
However, by comparing expressions~(\ref{eq:extended-tautological-form})
and (\ref{dm2}), one finds that $\Lambda$ can only be a function
of $q$ and $\tau$, i.e., $\Lambda=\Lambda\left(q,\tau\right)$.
As a result, we write $dM$ as 
\begin{equation}
dM=\varpi dq-\frac{B_{D-2}}{8\pi}\frac{q^{\frac{D-1}{D-2}}}{D-1}\frac{\partial\Lambda}{\partial\tau}d\tau\,,\label{eq:dm3}
\end{equation}
and we arrive at the additional constraint%
\footnote{Expression~(\ref{eq:chi}) is the analog of constraint~(9) in \cite{baldfresmol2016}.%
} 
\begin{equation}
\chi=\xi+\frac{1}{8\pi}\frac{B_{D-2}}{\left(D-1\right)}q^{\frac{D-1}{D-2}}\frac{\partial\Lambda}{\partial\tau}\,.\label{eq:chi}
\end{equation}
The equation $\chi=0$ reduces the tautological form $pdq+\xi d\tau$
in the extended phase space to the Poincar\'e-Cartan form $pdq-hd\tau$
in the reduced phase space, where 
\begin{equation}
h\left(q,\tau\right)=\frac{1}{8\pi}\frac{B_{D-2}}{\left(D-1\right)}q^{\frac{D-1}{D-2}}\frac{\partial\Lambda}{\partial\tau}~.\label{h2}
\end{equation}
The appropriate symplectic form in the extended phase space is given
by the extension of~(\ref{eq:reduced-symplectic}) which preserves
the canonical relations between the coordinates $\eta$, 
\begin{equation}
\tilde{\omega}=\left(\frac{\partial\varpi}{\partial p}\right)dq\wedge dp+\left(\frac{\partial\varpi}{\partial\tau}\right)dq\wedge d\tau+d\tau\wedge d\xi\,.
\end{equation}
In the transformed coordinates $\eta$, the symplectic form $\tilde{\omega}$
gives rise to the canonical Poisson brackets 
\begin{equation}
\left\{ f\left(\eta\right),g\left(\eta\right)\right\} =\frac{\partial f}{\partial q}\frac{\partial g}{\partial\varpi}+\frac{\partial f}{\partial\tau}\frac{\partial g}{\partial\xi}-f\leftrightarrow g\,.\label{eq:poisson-brackets}
\end{equation}
With respect to the Poisson brackets in Eq.~(\ref{eq:poisson-brackets}),
the constraints are first-class, so the number of physical degrees
of freedom is zero, as dictated by the general theory \cite{baldfresmol2016}.

Given that $M$ has been defined to be a thermodynamic potential,
one has $\varpi dq\equiv T_{\mathrm{eff}}dS$, where the thermodynamic
temperature $T_{\mathrm{eff}}$ is given by Eq.~(\ref{eq:varpi}),
\begin{equation}
T_{\mathrm{eff}}=T-\frac{1}{8\pi}\frac{B_{D-2}}{\left(D-1\right)}\left(\frac{4S}{B_{D-2}}\right)^{\frac{D-1}{D-2}}\frac{\partial\Lambda}{\partial S}\,.\label{ttermo}
\end{equation}
The identification of $-\tau$ with the thermodynamic pressure $P$
leads to $dM=T_{\mathrm{eff}}dS$ being the exchanged heat in an isobaric
process, which is the enthalpy.%
\footnote{We note that the identifications made between mechanical and thermodynamic
variables are not fundamental, serving mainly to make contact with
previous works. The only physically meaningful statements are that
$M$ is a thermodynamic potential and that $T_{\mathrm{eff}}$ is
the integrating factor for entropy (that is, the temperature).%
} From the constraint~(\ref{eq:chi}) one has 
\begin{equation}
V=-\frac{1}{8\pi}\frac{B_{D-2}}{\left(D-1\right)}\left(\frac{4S}{B_{D-2}}\right)^{\frac{D-1}{D-2}}\frac{\partial\Lambda}{\partial P}\,,\label{eq:volume}
\end{equation}
where the conjugate variable $\xi$ is the thermodynamic volume.

\subsection{Connection with the Schwarzschild solution}

In order to make the connection with the Schwarzschild solution, we
rewrite $dM$ in Eq.~(\ref{dm2}) as 
\begin{equation}
dM=\varpi^{\prime}dq-\frac{1}{8\pi}\frac{B_{D-2}}{D-1}d\left(q^{\frac{D-1}{D-2}}\Lambda\right)\,,\,\,\varpi^{\prime}=\frac{B_{D-2}}{4\pi}\left(p+\frac{1}{2}\frac{q^{\frac{1}{D-2}}\Lambda}{D-2}\right)\,.
\end{equation}
From the general theory presented in \cite{baldfresmol2016}, this
identity implies that there is a time-dependent (i.e., $\tau$-dependent)
canonical transformation $q\mapsto q$, $\varpi\mapsto\varpi^{\prime}$
mapping the reduced phase space with coordinates $\left(\varpi,q\right)$
to a reduced phase space with coordinates $\left(\varpi^{\prime},q^{\prime}\right)$.
The associated tautological form is $dM^{\prime}=\varpi^{\prime}dq^{\prime}$,
with 
\begin{equation}
M^{\prime}=M-E_{\Lambda}\,,\ E_{\Lambda}=\theta\frac{\Lambda}{8\pi}~,\label{m}
\end{equation}
and $\theta$ given in Eq.~(\ref{vol}).

In fact, consider the (second type) generating function 
\begin{equation}
F_{\Lambda}\left(\varpi^{\prime},q,\tau\right)=-\frac{B_{D-2}}{8\pi}\frac{1}{D-1}q^{\frac{D-1}{D-2}}\Lambda+\varpi^{\prime}q\label{tc}
\end{equation}
for the transformation $\left(\varpi,q\right)\mapsto\left(\varpi^{\prime},q^{\prime}\right)$.
One has 
\begin{align}
\varpi & =\frac{\partial F_{\Lambda}}{\partial q}=-\frac{\Lambda}{8\pi}\frac{B_{D-2}}{D-2}q^{\frac{1}{D-2}}-\frac{1}{8\pi}\frac{B_{D-2}}{D-1}q^{\frac{D-1}{D-2}}\frac{\partial\Lambda}{\partial q}+\varpi^{\prime}\nonumber \\
 & =\frac{B_{D-2}}{4\pi}\left(p-\frac{1}{2}\frac{1}{D-1}q^{\frac{D-1}{D-2}}\frac{\partial\Lambda}{\partial q}\right)
\end{align}
which is precisely the result presented in Eq.~(\ref{eq:varpi}).
Moreover, 
\begin{equation}
q^{\prime}=\frac{\partial F_{\Lambda}}{\partial\varpi^{\prime}}=q\,,\,\, h^{\prime}=h+\frac{\partial F_{\Lambda}}{\partial\tau}\equiv0\,.\label{qh}
\end{equation}
Transformations of the form $F_{\Lambda}\left(\varpi^{\prime},q,\tau\right)=g\left(q,\tau\right)+\varpi^{\prime}q$
guarantee that $q^{\prime}=q$ and preserve the Bekenstein's notion
of entropy and the second law of black hole Thermodynamics.

We note that $\Lambda$ is not a thermodynamic variable, but a parametrizing
function for all extended phase spaces, all of which are in the same
equivalence class of the reduced phase space with coordinates $\left(\varpi^{\prime},q^{\prime}\right)$
modulo time-dependent canonical transformations. In this sense, one
really has a family of constraints $\chi\left(\Lambda\right)$ and
$\phi\left(\Lambda\right)$. On the other hand, for each choice of
$\Lambda$, the constraints $\chi\left(\Lambda\right)$ and $\phi\left(\Lambda\right)$
lead to different equations of state~(\ref{ttermo}) and (\ref{eq:volume}),
so to physically different thermodynamic systems. The case where $\Lambda$
is constant corresponds to the minimal description presented in section~\ref{sub:Minimal-Thermodynamics-for}.
The case $\Lambda=8\pi\tau$ corresponds to the introduction of a
thermodynamic volume in section \ref{first_extension}.

In the primed coordinate system $\left(\varpi^{\prime},q^{\prime}\right)$,
constraint~(\ref{eq:phi-constraint}) and $M^{\prime}$ become 
\begin{align}
 & \phi^{\prime}=\frac{4\pi}{B_{D-2}}\varpi^{\prime}-\frac{1}{4}\left(D-3\right)q^{\prime-\frac{1}{D-2}}~,\nonumber \\
 & M^{\prime}=\frac{B_{D-2}}{16\pi}\left(D-2\right)q^{\prime\frac{D-3}{D-2}}~,~q^{\prime}=q=\frac{4S}{B_{D-2}}~.\label{m-linha}
\end{align}
The quantity $M^{\prime}$ in Eq.~(\ref{m-linha}) is the mass of
the $D$-dimensional Schwarzschild black hole, described by the metric~(\ref{metric})
with $\Lambda=0$.%
\footnote{It is clear that in the primed system the evolution parameter $\tau$
plays no role, since, from Eq.~(\ref{qh}), $dM^{\prime}=\varpi^{\prime}dq^{\prime}-h^{\prime}d\tau\equiv\varpi^{\prime}dq^{\prime}$.%
}

One sees that the reduced phase space case corresponds to the asymptotically
flat case. Furthermore, $\Lambda/8\pi$ is the spacetime energy per
unit volume \cite{Joh14} and $-\theta$ is the volume of the sphere
of radius $r_{+}$. Thus, $E_{\Lambda}$ in Eq.~(\ref{m}) can be
interpreted as the energy removed from spacetime due to the presence
of the black hole.

The form of the transformation $F_{\Lambda}$ can be inferred by noting
that all the effects due to a negative cosmological constant $\Lambda$
are included into the problem by adding the energy $E_{\Lambda}$
to the Schwarzschild mass $M^{\prime}$. Therefore, all thermodynamic
results commented in section~\ref{sec:AdS-black-holes} can be obtained
from the Schwarzschild solution (primed system) without ever having
to solve the Einstein equations, just by using the canonical transformation
(\ref{tc}). In particular, we use the canonical transformation~(\ref{tc})
to obtain the expressions in section~\ref{first_extension}, i.e.,
the Smarr formula for $\Lambda\neq0$. This is possible, once the
expression (\ref{m-linha}) for $M^{\prime}$ (as well as the pure
Schwarzschild case) is homogeneous in $q^{\prime}$, unlike the SAdS
case. Euler's theorem for homogeneous functions \cite{kastor2009}
states that, if $G\left(\lambda^{\alpha_{1}}x_{1},\lambda^{\alpha_{2}}x_{2}\right)=\lambda^{r}G\left(x_{1},x_{2}\right)$,
then 
\begin{equation}
rG=\alpha_{1}\left(\frac{\partial G}{\partial x_{1}}\right)x_{1}+\alpha_{2}\left(\frac{\partial G}{\partial x_{2}}\right)x_{2}~.\label{hom}
\end{equation}
Thus, Euler's formula~(\ref{hom}) with $G=M^{\prime}$, $x_{1}=q$
and $x_{2}=0$ gives 
\begin{equation}
\left(D-3\right)M^{\prime}=\left(D-2\right)\varpi^{\prime}q^{\prime}~.
\end{equation}
Then, using Eqs.~(\ref{m-linha}) and (\ref{qh}), we obtain 
\begin{equation}
\left(D-3\right)M=\frac{B_{D-2}}{4\pi}\left(D-2\right)pq+\frac{B_{D-2}}{4\pi}\frac{\Lambda}{\left(D-1\right)}q^{\frac{D-1}{D-2}}~.\label{s1}
\end{equation}

The next step is to write the expression~(\ref{s1}) in terms of
the thermodynamic variables ($\tau=-P$ and $\xi=V$). From $dM=pdq-hd\tau$
and Eq.~(\ref{h2}), 
\begin{equation}
h=\frac{1}{8\pi}\frac{B_{D-2}}{\left(D-1\right)}q^{\frac{D-1}{D-2}}\frac{\partial\Lambda}{\partial\tau}=-\frac{\partial M}{\partial\tau}=-\frac{\partial M}{\partial\Lambda}\frac{\partial\Lambda}{\partial\tau}~.\label{h3}
\end{equation}
Using Eq.~(\ref{h3}), Eq.~(\ref{s1}) is expressed as 
\begin{equation}
\left(D-3\right)M=\left(D-2\right)TS-2\Lambda\frac{\partial M}{\partial\Lambda}~.\label{SAdS-smarr}
\end{equation}
The relation (\ref{SAdS-smarr}) is the Smarr formula~(\ref{smarr})
for SAdS in $D$ dimensions. As a result, the $D$-dimensional Smarr
relation~(\ref{SAdS-smarr}) is not the homogeneity relation in the
extended phase space, but the image of the Euler relation for the
Schwarzschild solution by the canonical transformation in Eq.~(\ref{tc}).

\section{\label{sec:The-Euler-relation}The Euler relation}

In this section we impose homogeneity of the equations of state, as
is required by a consistent black hole Thermodynamics \cite{bel2-2005,bel2005}.
Even though homogeneity does not completely fix the dependence
of $\Lambda$ on phase space, it imposes strong restrictions on the
functional form of $\Lambda\left(S,P\right)$.

Indeed, homogeneity gives us a multiplicity of interesting scenarios.
From the family of theories that can be generated by the canonical
transformation in Eq.~(\ref{tc}), we focus on the subset of possible
thermodynamic descriptions for the AdS black holes which respect the
homogeneity condition. This set is parametrized by phenomenological
constants, as we discuss in the following.

Homogeneity with respect to the rescaling $S\rightarrow\lambda S$
in~(\ref{eq:kappa}) leads to 
\begin{equation}
T\rightarrow\lambda^{-\frac{1}{D-2}}T\,\,,
\end{equation}
and 
\begin{equation}
\Lambda\rightarrow\lambda^{\frac{2}{2-D}}\Lambda~,V\rightarrow\lambda^{\frac{D-3}{D-2}-c}V~,\ P\rightarrow\lambda^{c}P~,\ M\rightarrow\lambda^{\frac{D-3}{D-2}}M~,
\end{equation}
where $c$ is an arbitrary constant. This constant $c$ parametrize\emph{s}
the set of homogeneous thermodynamic descriptions.

Applying Euler's formula~(\ref{hom}) with $G=M$, $x_{1}=S$ and
$x_{2}=P$, we get $\alpha_{1}=1$, $\alpha_{2}=c$, and $r=\left(D-3\right)/\left(D-2\right)$.
The Euler relation in the extended space becomes 
\begin{equation}
\left(D-3\right)M=\left(D-2\right)\left[T_{\mathrm{eff}}S+cVP\right]~.\label{re}
\end{equation}
Analogously, the homogeneity condition of the equation of state $\Lambda=\Lambda\left(S,P\right)$,
i.e., setting $G=\Lambda,$ $r=2/(2-D)$, $x_{1}=S$, $x_{2}=P$,
$\alpha_{1}=1$ and $\alpha_{2}=c$ in Eq.~(\ref{hom}), provides
\begin{equation}
\Lambda=\frac{D-2}{2}\left(-c\frac{\partial\Lambda}{\partial P}P-\frac{\partial\Lambda}{\partial S}S\right)~.\label{Lambda_S_P}
\end{equation}

For $c=0$, the expression for $\Lambda$ in Eq.~(\ref{Lambda_S_P})
gives 
\begin{equation}
\Lambda=S^{\frac{2}{2-D}}\tilde{f}\left(P\right)~,
\end{equation}
with $\tilde{f}$ an arbitrary nonvanishing function of $P$. In the
present case, homogeneity does not fix the dependency of $\Lambda$
on $P$. Moreover, let us consider the heat capacity at constant pressure $C_{P}$, with
\begin{equation}
C_{P}=-\pi T_{\mathrm{eff}}\frac{\left(D-2\right)}{\left(D-3\right)}\frac{B_{D-2}}{g_{D}\left(\Lambda\right)}\left(\frac{4S}{B_{D-2}}\right)^{\frac{D-1}{D-2}}~,
\end{equation}
where 
\begin{equation}
g_{D}\left(\Lambda\right)=1+\frac{2\left|\Lambda\right|}{\left(D-2\right)}\frac{1}{\left(D-1\right)}\left(\frac{4S}{B_{D-2}}\right)^{\frac{2}{D-2}}~.
\end{equation}
We observe that $g_{D}>0$, and consequently the system is always unstable. That is, the SAdS Thermodynamics with $c=0$
is similar to its Schwarzschild counterpart, in the sense that it cannot describe stability or phase transitions.

We now consider scenarios where $c\neq0$. The general solution of
Eq.~(\ref{Lambda_S_P}) can be written as 
\begin{equation}
\Lambda\left(S,P\right)=P^{\frac{1}{c}\frac{2}{2-D}}f\left(SP^{-\frac{1}{c}}\right)~,
\label{f}
\end{equation}
where $f$ is an arbitrary function.%
\footnote{It should be observed that $SP^{-\frac{1}{c}}$ is a homogeneous quantity
of zero order. Therefore, $\Lambda$ is homogeneous irrespective of
$f$.%
}
We fix $f$ demanding that the zeroth law of black hole Thermodynamics
is preserved.
For this purpose, using~(\ref{f}), we rewrite Eq.~(\ref{ttermo}) as
\begin{equation}
T_{\mathrm{eff}}=\frac{D-3}{4\pi}\left(  \frac{4S}{B_{D-2}}\right)  ^{\frac
{1}{2-D}}+\frac{\tilde{\Lambda}}{4\pi}\left[  \frac{2-D}{f}X\frac{df}%
{dX}-D+1\right]  \left(  \frac{4S}{B_{D-2}}\right)  ^{\frac{1}{D-2}%
}\,,\label{ttermo2}%
\end{equation}
with $X=SP^{-\frac{1}{c}}$. To guarantee consistency in the
association between the thermodynamic temperature $T_{\mathrm{eff}}$ and the
black hole surface gravity $\kappa$, we require that $T_{\mathrm{eff}}$ and
$\kappa$ should have the same functional dependence on $S$ and $\Lambda$, and
consequently that the expressions for $T_{\mathrm{eff}}$ and $\kappa$ must
have the same powers of $S$ and $\Lambda$. In addition, we demand that in the
limit $\Lambda\rightarrow0$ we should recover the asymptotically flat
expressions, that is, $T_{\mathrm{eff}}\rightarrow\kappa/2\pi$. 

Comparing Eqs.~(\ref{ttermo2}) and (\ref{kappa-1}), previous
considerations imply that the terms inside the brackets of Eq.~(\ref{ttermo2})
must be a constant, 
\begin{equation}
\frac{2-D}{f}X\frac{df}{dX}=2a~,\label{eq-f}%
\end{equation}
where $a$ is an arbitrary constant. Solving Eq.~(\ref{eq-f}) and using Eq.~(\ref{f}), we obtain
\begin{equation}
\Lambda = -K\left[  \left(  \frac{4S}{B_{D-2}}\right)  ^{a}P^{b/c}\right]
^{\frac{2}{2-D}}~,\ a+b=1~,\ 2b\neq\left(  2-D\right)  c ~,
\label{lamb}%
\end{equation}
with $K$ an integration constant (we set $K=1$ from this point on). 
For $2b=\left(2-D\right)c$, $V$ does not depend on $P$ and the
Legendre transformation $\left(S,V\right)\rightarrow\left(S,P\right)$
is singular. In Eq.~(\ref{ttermo}), we use Eq.~(\ref{lamb}) and
Eq.~(\ref{eq:momento}) to eliminate $T$ in order to obtain 
\begin{equation}
T_{\mathrm{eff}}=\frac{1}{4\pi}\left[\left(D-3\right)\left(\frac{4S}{B_{D-2}}\right)^{\frac{1}{2-D}}-\frac{2\Lambda\left(D-2a-1\right)}{\left(D-2\right)\left(D-1\right)}\left(\frac{4S}{B_{D-2}}\right)^{\frac{1}{D-2}}\right]~.\label{teff}
\end{equation}
We emphasize that  $T_{\mathrm{eff}}$ differs from the original Hawking temperature $T$ in Eq.~(\ref{usual}). It shows that the Hamiltonian approach used here can lead to distinctive physical results.

Also, we note that $T_{\mathrm{eff}}=\kappa_{\mathrm{eff}}/2\pi$,
where $\kappa_{\mathrm{eff}}$ is given by Eq.~(\ref{kappa-1}) with
the cosmological constant substituted by an effective version, 
\begin{equation}
\Lambda\rightarrow\Lambda_{\mathrm{eff}}=\left(1-\frac{2a}{D-1}\right)\Lambda~.
\end{equation}
Given the solution for $f$ from Eq.~(\ref{eq-f}),
the temperature
$T_{\mathrm{eff}}$ is proportional to an effective surface gravity
$\kappa_{\mathrm{eff}}$ of a SAdS background with effective cosmological
constant $\Lambda_{\mathrm{eff}}$. That is, the geometrical interpretation
of the black hole temperature remains valid, provided one assumes
that the spacetime has an effective cosmological constant. 

Substituting result~(\ref{lamb}) in expression (\ref{eq:volume})
for $V$ and in expression (\ref{teff}) for $T_{\mathrm{eff}}$,
we obtain the equations of state 
\begin{align}
 & P^{\gamma^{-1}}V=\frac{B_{D-2}}{8\pi}\frac{\left(1-\gamma^{-1}\right)}{D-1}\left(\frac{4S}{B_{D-2}}\right)^{\frac{D-1-2a}{D-2}}~,\ \gamma^{-1}=1+\frac{2b}{\left(D-2\right)c}~,\label{es}\\
 & T_{\mathrm{eff}}=\frac{1}{4\pi}\left(\frac{4S}{B_{D-2}}\right)^{\frac{1}{2-D}}\left[D-3+\frac{2\left(D-2a-1\right)}{\left(D-2\right)\left(D-1\right)}\left(4\frac{SP^{-\frac{1}{c}}}{B_{D-2}}\right)^{\frac{2b}{D-2}}\right]\,.\label{es2}
\end{align}
The new equation of state (\ref{es}) is a consequence of an additional
thermodynamic variable in the configuration space. Since our system
has two degrees of freedom, it follows that Eqs.~(\ref{es}) and
(\ref{es2}) completely characterize the Thermodynamics.

From Eq.~(\ref{es}) we see that $PV^{\gamma}$ is constant for an
isentropic process. Thus one can compare $\gamma$ with the isentropic
expansion factor of the ideal gas (where $\gamma=C_{P}/C_{V}$). In
the ideal gas case the factor $\gamma$ is associated with the degrees
of freedom of the microscopic description (e.g., vibration or rotation
modes). The equation of state~(\ref{es}) suggests that the constants
$b$ and $c$ are related to the microscopic features of the theory.

We can now investigate other properties of the black hole system,
such as its thermal capacity under constant pressure $C_{P}$, 
\begin{equation}
C_{P}=-\pi T_{\mathrm{eff}}\frac{\left(D-2\right)B_{D-2}}{f_{a,D}\left(\Lambda\right)}\left(\frac{4S}{B_{D-2}}\right)^{\frac{D-1}{D-2}}~,\label{cp}
\end{equation}
where 
\begin{equation}
f_{a,D}\left(\Lambda\right)=D-3-2\left|\Lambda\right|\frac{\left(1-2a\right)\left(D-1-2a\right)}{\left(D-2\right)\left(D-1\right)}\left(\frac{4S}{B_{D-2}}\right)^{\frac{2}{D-2}}~.
\end{equation}
A negative value for $C_{P}$ in Eq.~(\ref{cp}) implies thermodynamic
instability. If $1/2\leq a\leq a_{\mathrm{crit}}$, with 
\begin{equation}
a_{\mathrm{crit}}=\frac{D-1}{2}\,,\label{a_ext}
\end{equation}
the system is unstable for any value of $\Lambda$. For the critical
value $a=a_{\mathrm{crit}}$, $\Lambda_{\mathrm{eff}}=0$ and the
AdS black hole behaves as its Schwarzschild counterpart.

We stress that $a$ is a phenomenological parameter; hence, consistency
conditions alone do not fix $a$, and a microscopic theory is necessary.
For instance, a physical interpretation for instability in the threshold
$a\in\left[1/2,a_{\mathrm{crit}}\right]$ is a microscopic theory
whose boundary conditions are not reflexive. As it is well known,
asymptotically AdS geometries are not globally hyperbolic, and therefore
field theories in those spacetimes must be supplemented with additional
boundary conditions at spacial infinity. For the so-called reflexive
boundary conditions \cite{Avis,Ishibashi}, AdS spacetimes behave
as a ``box'' considering the energy content associated to the matter
field. The enforcement of reflexive boundary condition guarantees
that energy does not escape throughout the spacial infinity. We suggest
that, in those cases, the proper thermodynamic description has $a\notin\left[1/2,a_{\mathrm{crit}}\right]$.
On the other hand, if one allows energy to escape throughout spacial
infinity (as it happens in the usual Schwarzschild spacetime) the
Thermodynamics cannot be stable, and so $a\in\left[1/2,a_{\mathrm{crit}}\right]$.

For $a>a_{\mathrm{crit}}$, the second term inside the brackets in
Eq.~(\ref{teff}) becomes negative (since $\Lambda<0$). Hence the
condition $T_{\mathrm{eff}}\geq0$ implies 
\begin{equation}
\left\vert \Lambda\right\vert \leq\left\vert \Lambda_{\mathrm{ext}}\right\vert ~,\ \Lambda_{\mathrm{ext}}=\frac{\left(D-3\right)\left(D-2\right)\left(D-1\right)}{2\left(D-1-2a\right)}\left(\frac{B_{D-2}}{4S}\right)^{\frac{2}{D-2}}~.\label{l-ext}
\end{equation}
In that case, we observe an extreme value for the cosmological constant,
$\Lambda=\Lambda_{\mathrm{ext}}$. Extreme black holes have a null
temperature $T_{\mathrm{eff}}$ with a non-null entropy $S$. Therefore,
it is possible that a SAdS black hole be extreme (in the thermodynamic
sense) even when it is not extreme in the geometric sense (the surface
gravity is non vanishing). This property might lead to several features
that are only expected in more complex scenarios where electric charge
or angular momentum is taken into account \cite{Dol1,KubMan2012,Deb16}.
In addition, from the expression for the thermal capacity~(\ref{cp}),
we observe that if $a\notin\left[1/2,a_{\mathrm{crit}}\right]$ the
system is stable for $f_{a,D}<0$, that is, 
\begin{equation}
\left|\Lambda\right|>\frac{\left(D-3\right)\left(D-2\right)\left(D-1\right)}{2\left(1-2a\right)\left(D-1-2a\right)}\left(\frac{B_{D-2}}{4S}\right)^{\frac{2}{D-2}}~.\label{tf}
\end{equation}

If $D=4$ the parameter $a$ can be fixed if one demands that the
black hole has the behavior predicted by Hawking-Page theory \cite{Haw1983}.
In that approach, spacetime is stable for $\left|\Lambda\right|>\pi/S$. Combining the relation $\left|\Lambda\right|>\pi/S$ with
the inequality in~(\ref{tf}) for $D=4$, we see that stability occurs for $a=0$ and for $a=2$.

For $a=0$ no thermodynamic quantities depend on the choice of $M=H$
nor $M=U$. Furthermore, if we set $c=-1$, we recover the Thermodynamics discussed in section~\ref{first_extension}. However, as 
presented
in Eq.~(\ref{lamb}), the case $a=0$ is consistent only if $c\neq-1$.
Despite this fact, this scenario appears to be the only one considered
in the literature so far.

One interesting (and so far unexplored) particular case is obtained
considering $a=2$, where 
\begin{equation}
T_{\mathrm{eff}}=\frac{1}{4\pi}\left[\left(\frac{S}{\pi}\right)^{-1/2}+\frac{\Lambda}{3}\left(\frac{S}{\pi}\right)^{1/2}\right]~,\ 
\Lambda = - \frac{\pi^{2} P^{1/c}}{S^{2}}~,\ c \notin\{0,1\}~.
\end{equation}
In this scenario, Thermodynamics has the instability-stability transition predicted by Hawking and Page.
But due to the presence of the extra thermodynamic
variable, the black hole has a temperature which is
not given by Hawking's formula~(\ref{usual}). Moreover, conditions~(\ref{tf}) and (\ref{l-ext})
imply that $\pi/S<\left\vert \Lambda\right\vert \leq3\pi/S$, which
can be translated, with Eq.~(\ref{eq:M}), to a relation between
mass and entropy: 
\begin{equation}
\frac{2}{3}<M\sqrt{\frac{\pi}{S}}\leq1~.
\end{equation}
The upper limit is always valid, while the lower limit is valid when
the system is stable.

\section{\label{sec:Conclusions}Conclusions and future perspectives}

We have considered the Thermodynamics of Schwarzschild-anti de Sitter
black holes. We extend the minimal thermodynamic setup, briefly reviewed
in section~\ref{sub:Minimal-Thermodynamics-for}, using the Hamiltonian
approach. In this formalism, thermodynamic equations of state are
realized as constraints on phase space. We demonstrate that the cosmological
constant $\Lambda$ can be introduced in the thermodynamic description
of the SAdS black hole as a result of a canonical transformation of
the Schwarzschild problem, $\Lambda$ being closely related to a generalized
thermodynamic volume. Our proposal differs from the ones already presented
in the literature (for example in \cite{21,seki2006,29}) since it
does not consider $\Lambda$ as a new thermodynamic variable. In fact,
using Legendre transformations, it is not possible to construct a
fundamental equation where the cosmological constant is an independent
variable. This is an essential point for the consistency of the thermodynamic
description.

It should be stressed that if the cosmological constant is not part
of the thermodynamic description, the equations of state are not homogeneous.
Homogeneity is required for extensivity \cite{kastor2009} as well
for the existence of an integrating factor for the reversible heat
exchange and, consequently, for the well-definition of the entropy
\cite{bel2-2005}. Therefore, even though the minimal approach is
consistent as far as thermodynamic potentials are required, it does
not provide the Euler relation. 
As discussed, previous treatments that include the
cosmological constant as a thermodynamic variable are not consistent. 
Our approach addresses both homogeneity and
the cosmological constant problems. In addition, we show that the SAdS
black hole can have an extreme behavior and can possess critical points.

Some degree of arbitrariness in the development is unavoidable, given
the great generality of the formalism used \cite{baldfresmol2016}.
This arbitrariness is related to the plethora of microscopic theories
that are compatible with our macroscopic treatment. Indeed, our treatment
has parameters that can be associated to different kinds of AdS boundary
conditions. In effect, it is natural that the behavior of an underlying
microscopic theory at the AdS boundary should have an important role
in how energy is stored (or lost) in the AdS spacetime, and therefore
have an effect in the Thermodynamics and in the stability conditions
associated to this geometry.

Nonetheless, thermodynamic considerations alone are an important guide
to fix the macroscopic theory. For instance, 
the development leading to the function $\Lambda(S,P)$ in Eq.~(\ref{lamb})
guarantees that the product $PV$ is proportional to $\theta \Lambda$. In this way, we are not only preserving the zeroth law, but also ensuring that our treatment is compatible with previous works and maintains the usual interpretation of $PV$ as the energy removed from spacetime due to the presence of the black hole \cite{Joh14}.

Concerning the physical characterization of the new variables $\tau$ 
and $\xi$ introduced in section~\ref{sec:A-Symplectic-approach}, 
there is no need to identify them with the thermodynamic pressure and 
volume, respectively, as suggested in Eq.~(\ref{eq:volume}). 
They can be interpreted as an arbitrary conjugate pair, with the parameter $M$ 
seen as an arbitrary thermodynamic potential. 
In this broader view, if the (chargeless and spinless) black hole has any free parameter other than its mass, it follows that its temperature will be given by Eq.~(\ref{teff}). On the other hand, if mass is the only free parameter, its associated temperature is given by the usual Hawking's formula~(\ref{usual}). 
In a sense, characterization of the thermodynamic quantities are related to the number of degrees of freedom of the system.

In this work, although we considered $\Lambda$ to be a function on
phase space, we assumed $\Lambda$ is constant on the spacetime manifold.
But it is worth pointing out that the Thermodynamics
we have developed is still consistent if $\Lambda$ varies in spacetime.
In that case, we would not be dealing with the Schwarzschild-anti
de Sitter solution of the Einstein's equations, but with a scenario
where $\Lambda$ is a dynamic geometric quantity \cite{kastor2009}.
And when extended to the sector of positive cosmological constant,
the treatment developed with a dynamical $\Lambda$ could have applications
in cosmology \cite{Lima}. We also anticipate that the formalism
presented here could be adapted to systems that have new degrees of
freedom such as the scenarios where soft-symmetries are relevant \cite{Haw2016}.


\begin{acknowledgments}

C. M. is supported by FAPESP, Brazil [grant number 2015/24380-2]; and CNPq,
Brazil [grant number \linebreak 307709/2015-9].

\end{acknowledgments}

\bigskip{}


\begin{thebibliography}{99}

\bibitem{maldacena}J. M. Maldacena, The Large N limit of superconformal
field theories and supergravity, Adv. Theor. Math. Phys. \textbf{2}
(1998) 231. arXiv:hep-th/9711200.

\bibitem{witten}E. Witten, Anti-de Sitter space and holography, Adv.
Theor. Math. Phys. \textbf{2} (1998) 253. arXiv:hep-th/9802150.

\bibitem{gubser}S. S. Gubser, Igor R. Klebanov, Alexander M. Polyakov,
Gauge theory correlators from noncritical string theory, Phys. Lett.
B \textbf{428} (1998) 105. arXiv:hep-th/9802109.

\bibitem{brown}J. D. Brown, J. Creighton, Robert B. Mann, Temperature,
energy and heat capacity of asymptotically anti-de Sitter black holes,
Phys. Rev. D \textbf{50} (1994) 6394. arXiv:gr-qc/9405007.

\bibitem{louko}J. Louko, S. N. Winters-Hilt, Hamiltonian thermodynamics
of the Reissner-Nordstrom anti-de Sitter black hole, Phys. Rev. D
\textbf{54} (1996) 2647. arXiv:gr-qc/9602003.

\bibitem{hemming}S. Hemming, L. Thorlacius, Thermodynamics of Large
AdS Black Holes, JHEP \textbf{0711} (2007) 086. arXiv:0709.3738.

\bibitem{grumiller}D. Grumiller, R. McNees, Thermodynamics of black
holes in two (and higher) dimensions, JHEP \textbf{0704} (2007) 074.
arXiv:hep-th/0703230.

\bibitem{rajeev}S. G. Rajeev, A Hamilton-Jacobi formalism for thermodynamics,
Ann. Phys. \textbf{323} (2008) 2265. arXiv:0711.4319.

\bibitem{morgan}J. Morgan, V. Cardoso, A. S. Miranda, C. Molina,
V. T. Zanchin, Gravitational quasinormal modes of AdS black branes
in d spacetime dimensions, JHEP \textbf{0909} (2009) 117. arXiv:0907.5011.

\bibitem{miranda}A. S. Miranda, C. A. B. Bayona, H. Boschi-Filho,
N. R. F. Braga, Black-hole quasinormal modes and scalar glueballs
in a finite-temperature AdS/QCD model, JHEP \textbf{0911} (2009) 119.
arXiv:0909.1790.

\bibitem{hubeny}V. E. Hubeny, D. Marolf, M. Rangamani, Hawking radiation
from AdS black holes, Class. Quant. Grav. \textbf{27} (2010) 095018.
arXiv:0911.4144.

\bibitem{Rabin}R. Banerjee, S. K. Modak, D. Roychowdhury, A unified
picture of phase transition: from liquid-vapour systems to AdS black
holes, JHEP \textbf{1210} (2012) 125. arXiv:1106.3877.

\bibitem{menoufi}B. M. El-Menoufi, B. Ett, D. Kastor, J. Traschen,
Gravitational Tension and Thermodynamics of Planar AdS Spacetimes,
Class. Quant. Grav. \textbf{30} (2013) 155003. arXiv:1302.6980.

\bibitem{cardoso}V. Cardoso, \'{O}. J. C. Dias, G. S. Hartnett, L. Lehner,
J. E. Santos, Holographic thermalization, quasinormal modes and superradiance
in Kerr-AdS, JHEP \textbf{1404} (2014) 183. arXiv:1312.5323.

\bibitem{myung}Y. S. Myung, T. Moon, Thermodynamic and classical
instability of AdS black holes in fourth-order gravity, JHEP \textbf{1404}
(2014) 058. arXiv:1311.6985.

\bibitem{lemos}J. P. S. Lemos, F. J. Lopes, M. Minamitsuji, J. V.
Rocha, Thermodynamics of rotating thin shells in the BTZ spacetime,
Phys. Rev. D \textbf{92} (2015) 064012. arXiv:1508.03642.

\bibitem{wald}R. M. Wald, The Thermodynamics of Black Holes, Living
Rev. Relativity \textbf{4} (2001) 6. arXiv:gr-qc/9912119.

\bibitem{kastor2009}D. Kastor, S. Ray, J. Traschen, Enthalpy and
the Mechanics of AdS Black Holes, Class. Quant. Grav. \textbf{26}
(2009) 195011. arXiv:0904.2765.

\bibitem{Joh14}C. V. Johnson, Holographic Heat Engines, Class. Quant.
Grav. \textbf{31} (2014) 205002. arXiv:1404.5982.

\bibitem{nat2015}M. Natsuume, AdS/CFT Duality User Guide, Springer,
Japan (2015). arXiv:1409.3575.

\bibitem{21}C. Teitelboim, The cosmological constant as a thermodynamic
black hole parameter, Phys. Lett. B \textbf{158} (1985) 293.

\bibitem{seki2006}Y. Sekiwa, Thermodynamics of de Sitter Black Holes:
Thermal Cosmological Constant, Phys. Rev. D \textbf{73} (2006) 084009.
arXiv:hep-th/0602269.

\bibitem{29}M. Urano, A. Tomimatsu, and H. Saida, Mechanical First
Law of Black Hole Spacetimes with Cosmological Constant and Its Application
to Schwarzschild-de Sitter Spacetime, Class. Quantum Grav. \textbf{26}
(2009) 105010. arXiv:gr-qc/0903.4230.

\bibitem{Dol1}B. P. Dolan, Where Is the PdV in the First Law of Black
Hole Thermodynamics?, Open Questions in Cosmology, Gonzalo J. Olmo
(Ed.) (2012). arXiv:1209.1272.

\bibitem{2011}B. P. Dolan, Pressure and volume in the first law of
black hole thermodynamics, Class. Quantum Grav. \textbf{28} (2011)
235017. arXiv:1106.6260.

\bibitem{KubMan2012}D. Kubiznak, R. B. Mann, P-V criticality of charged
AdS black holes, JHEP \textbf{07} (2012) 033. arxiv:1205.0559.

\bibitem{baldfresmol2016}M. C. Baldiotti, R. Fresneda, C. Molina,
A Hamiltonian approach to Thermodynamics, Ann. Phys. \textbf{373}
(2016) 245. arXiv:1604.03117.

\bibitem{tangherlini}F. R. Tangherlini, Schwarzschild field in n
dimensions and the dimensionality of space problem, Nuovo Cim. \textbf{27}
(1963) 636.

\bibitem{Boucher}W. Boucher, G. W. Gibbons, and G. T. Horowitz, Uniqueness
theorem for anti-de Sitter spacetime, Phys. Rev. D \textbf{30} (1984)
2447.

\bibitem{bel2-2005}F. Belgiorno, Quasi-Homogeneous Thermodynamics
and Black Holes, J. Math. Phys. \textbf{44} (2003) 1089. arXiv:gr-qc/0210021.

\bibitem{smarr73}L. Smarr, Mass Formula For Kerr Black Holes, Phys.
Rev. Lett. \textbf{30} (1973) 71 {[}Erratum ibid \textbf{30} (1973)
521{]}.

\bibitem{kastor2010}D. Kastor, S. Ray, J. Traschen, Smarr Formula
and an Extended First Law for Lovelock Gravity, Class. Quant. Grav.
\textbf{27} (2010) 235014. arXiv:1005.5053.

\bibitem{bel2005}F. Belgiorno, Black Hole Thermodynamics in Carath\'{e}odory's
Approach, Phys. Lett. A \textbf{312} (2003) 324. arXiv:gr-qc/0210020.

\bibitem{Avis}S. J. Avis, C. J. Isham, and D. Storey, Quantum field
theory in anti-de Sitter space-time, Phys. Rev. D \textbf{18} (1978)
3565.

\bibitem{Ishibashi}A. Ishibashi, R. M. Wald, Dynamics in Non-Globally-Hyperbolic
Static Spacetimes III: Anti-de Sitter Spacetime, Class. Quant. Grav.
\textbf{21} (2004) 2981. arXiv:hep-th/0402184.

\bibitem{Deb16}U. Debnath, Entropy bound of horizons for accelerating,
rotating and charged Plebanski--Demianski black hole, Ann. Phys. \textbf{372}
(2016) 449. arXiv:1507.00901.

\bibitem{Haw1983}S. W. Hawking and D. N. Page, Thermodynamics Of
Black Holes In Anti-De Sitter Space, Commun. Math. Phys. \textbf{87}
(1983) 577.

\bibitem{Lima}J. A. S. Lima, S. Basilakos, J. Sol\`{a}, Thermodynamical
aspects of running vacuum models, Eur. Phys. J. C \textbf{76} (2016)
228. arXiv:1509.00163.

\bibitem{Haw2016}S. W. Hawking, M. J. Perry, and A. Strominger, Soft
Hair on Black Holes, Phys. Rev. Lett. \textbf{116} (2016) 231301.
arXiv:1601.00921. \end{thebibliography}
\end{document}